\documentclass[multphys,vecphys]{svmult}

\usepackage{makeidx}         
\usepackage{graphicx}        
\usepackage{multicol}        
\usepackage[bottom]{footmisc}

\makeindex             

\begin{document}

\title*{Probing $\mu$-arcsec astrometry with NACO}
\author{Andreas Seifahrt\inst{1,2} \and Tristan R\"oll\inst{2} \and Ralph Neuh\"auser\inst{2}}
\authorrunning{Seifahrt et al.}
\institute{European Southern Observatory (ESO), Garching, Germany
\texttt{aseifahr@eso.org}
\and Astrophysikalisches Institut und Universit\"ats-Sternwarte (AIU), Jena, Germany }
%
%
\maketitle

Relative astrometric measurements with a precision far better than 1 mas (milli-arcsec) are commonly
regarded as the domain of interferometry. Pioneering work by Pravdo \& Shaklan (1996), made in the 
optical, reached a precision of 150\,$\mu$arcsec in direct imaging but is ultimately limited by 
atmospheric turbulence and differential chromatic refraction (DCR) effects\index{differential chromatic refraction}. 
Neuh\"auser et al. (2006, 2007) demonstrated that AO assisted observations with NACO\index{NACO} in a near-infrared 
narrow band filter allow measurements with a precision of $\sim$50\,$\mu$as (micro-arsec) on a 0.6 arcsec binary 
within one hour and are unaffected by DCR effects. This opens new possibilities for astrometric detections 
of extrasolar planets and the determination of their true masses. We discuss here how to improve the 
measurements and address the necessary calibrations. 

\section{Motivation and Introduction}\label{sec:1}
The search for extrasolar planets\index{extrasolar planets} and their physical characterization has become 
an important field in today's astronomy and astrophysics. Dedicated instruments, such as HARPS on the ESO 
3.6m telescope on La Silla were designed for precise radial velocity (RV) measurements to find the 
fingerprints of extrasolar planets, imprinted in the RV signal of their host stars. With this method more 
than 200 extrasolar planet candidates have been found.\newline 
\textit{Candidates} -- because their masses are determind only to a lower limit, $m\sin{i}$. The 
inclination angle of the extrasolar planets are not constrained from RV measurements and the true mass 
of the planets is unknown. 

Only two methods can add enough information to allow the determination of the inclination angle and thus
the true mass of an extrasolar planet. In case of a nearly edge-on viewing geometry the planet can be
seen in the lightcurve of the host star as a transit. However, such events are rare. Only astrometric 
measurements of the induced positional \textit{wobble} of the host star can solve for the inclination 
of the system in all other cases. 

Moreover, searches for extrasolar planets by astrometry are sensitive to planets in wide orbits, in contrast to 
the RV technique. A number of target classes unsuitable for RV measurements, such as early-type stars and 
ultra-cool dwarfs, as well as fast rotating and otherwise active stars are also unproblematic for astrometry. 

Astrometric measurements have the only intrinsic limitation that a large number of suitable reference stars 
have to be in the field of view (FOV) to define a local restframe. This restricts observation to 
fields of high stellar density, e.g. to low galactic latitudes. In addition the reference stars itself have 
to have known proper motion and parallaxes. Moreover, chromatic refraction effects as well as atmospheric 
turbulence limits the achievable precision. Thus, the only successful observations of the astrometric 
wobble of exoplanet host stars have been achieved from space, namely for GJ 876 b\index{GJ 876 b}, 
55 Cancri d\index{55 Cancri d} and $\epsilon$ Eridani b\index{eps Eridani b}, using the Fine Guiding Sensor (FGS) 
of the Hubble Space Telescope, see e.g. Benedict et al. (2006) 

Neuh\"auser et al. (2006, 2007) proposed the use of adaptive optics (AO) in the near infrared to (a) work
in a regime where DCR effects are much smaller than in the optical and (b) to suppress the atmospheric turbulence.
Since the usable field of view of AO assisted measurements is limited to the isoplanatic angle, measurements have
to concentrate on physical binaries or multiple systems where the astrometric wobble of one component is
measured relative to the other component(s). With this approach the problem of relative motion of the 
reference stars is solved since parallax and proper motion are identical for both components, leaving 
only the orbital motion as an open parameter. 
First measurements, conducted with NACO on the VLT in December 2004 and October 2006 let to unprecedented 
precision in ground based astrometry.

\section{Results from the feasibility study}\label{sec:2}
A first feasibility study, started in 2004, concentrated on the binary systems HD19994\index{HD19994} and HD19063.\index{HD19063} The brighter component in HD19994 has a known exoplanet candidate as RV measurements 
by Mayor et al. (2004) revealed. The expected astrometric signal is at least 131 $\mu$as, when assuming the 
minimum mass $m\sin{i}$ as true mass and adopting the reported excentricity of the orbit.

Astrometric measurements with NACO were conducted in December 2004. The technique for high precision
astrometry is based on a principle used in RV measurements. If the resolution and sampling of a measurement 
is not sufficient, statistics over many independent measurements have to be used to reach the necessary 
precision. In the RV technique, measurements over many hundred spectral lines provide the necessary statistical 
basis. For astrometric measurements a high number of images have to be taken. Separation measurements in each 
individual frame are checked for a Gaussian distribution and the error of the mean separation can be computed 
from the standard deviation of the mean divided by the square root of the number of frames. From 120 frames of 
HD19994 and 60 frames of HD19063 that passed the statistical tests, a precision in the separation of the binary 
components of $\sim$92\,$\mu$as and $\sim$50\,$\mu$as, respectively, was reached within one hour. This
marks the best relative astrometric precision ever achieved with a single aperture telescope from the ground.

\section{Improvements: NACO cube mode}\label{sec:3}
Each individual frame was taken through a narrow-band filter and the exposure time was set to the minimum
DIT. Hence the measurements in 2004 where fully overhead dominated. Narrow band images with the fine
pixel scale of NACOs S13 camera are read-noise limited up to several minutes of exposure time. Hence no jitter
observations are necessary for sky subtraction and frames can be taken in staring mode. Thus, we used the cube 
mode of NACO in our second observation campaign in October 2006 to obtain more frames per given time than in 
autojitter mode. The chip had to be windowed to handle the high data rate. The readout overheads are strongly 
suppressed and several thousand frames can be taken per hour. Compared to the results from the previous run, 
we could improve our efficiency by more than a factor of 50.
 
\section{Calibration Issues}
The precision achieved here marks a breakthrough in ground based relative astrometry. It demonstrates 
that measurements in the separation of a close binary with a precision to about 4/100,000 are possible with NACO. 
This raises the question on the calibration of such measurements. In the scheme presented here, we do not 
need to provide an absolute calibration of the pixel scale. We are aiming to determine relative changes 
in the separation of a stellar binary of about 1/100 of a pixel, but not its absolute value. Thus, the uncertainties 
in the determination of the absolute pixel scale, typically of the order of 4/1000 (see Neu\"auser et al. 2006, or 
Chauvin et al. 2004) are not the dominating error source, especially since we are not aiming for high accuracy but
for high precision.
Instead we have to assure the stability of the pixel scale (or reversely the $f$-ratio of the imager) to 
better than 4/100,000. The typical calibration sources, HIPPARCOS binaries\index{Hipparcos binaries}, have precise coordinates but the uncertainties in their proper motion multiplied by the time since the HIPPARCOS 
epoch of 1991.25 rule out these sources as calibrators on the needed level of precision. The same holds 
for the SiO masers in the galactic center. Eventhough these objects have possibly the best known 
astrometric properties (from VLBA measurements), they are not fitting into the FOV of the S13 camera.

Hence, we have to look for an intrinsically stable reference system that would fit into the FOV of the S13 
camera. After a concise literature study, we selected 47 Tuc \index{47 Tuc} as a reference system. 
47 Tuc is an old globular cluster with a known and small velocity dispersion. McLaughlin et al. (2006) 
recently determined the two dimensional velocity dispersion in a spherical region around the cluster core 
within a 20 arcsec radius. For the red giant stars a value of $\sigma_\mu = 0.631^{+0.020}
_{-0.025}$ mas/year was reported. To determine the intrinsic astrometric stability, we have measured all 
separations between any two stars in one sub-field of 47 Tuc for the first time in October 2006 over a large 
number of frames and we will monitor these quantities in the next observing campaigns. 

The relative stability of the pixel scale, obtained with measurements of 47 Tuc depends on the number of 
stars in the observed sub-field and their mean separation. Due to error statistics, the standard deviation 
of the separation measurements can be divided by the square root of the number of independent separations,
which is $n - 1$ with the number $n$ of stars. This relation has been confirmed by a Monte Carlo simulation, 
tested on a Gaussian distributed field of 20 stars as well as on our true measurements that have the same
field density. Adopting a random velocity dispersion of 630 mas/year (hence, no symmetry), the stability 
of the pixel scale can be assured with a precision of $\sim$ 60$\mu$as for one year. Hence, the intrinsic 
relative stability of the reference system is mainly limited by the velocity dispersion and the number and 
geometry of the stars within the sub-field.

We conclude that high resolution astrometry with NACO is neither limited by the atmosphere nor by the 
instrument itself but by suitable reference objects with a high intrinsic stability. This is true as long 
as the stability of the pixel scale on a level of a few in hundred thousand can not be implied but has to 
be confirmed and monitored with on-sky measurements.

\printindex

\begin{thebibliography}{99.}
\bibitem{Benedict} Benedict, G.~F., et al.\ 2006, AJ, 132, 2206 
\bibitem{Chauvin04} Chauvin, G., Lagrange, A.-M., Dumas, C., Zuckerman, B., Mouillet, D., Song, I., Beuzit, J.-L., \& Lowrance, P.\ 2004, A\&A, 425, L29 
\bibitem{neuh07} Neuh\"auser, R., Seifahrt, A., Roell, T., Bedalov, A., \& Mugrauer, M.\ 2007, MNRAS, submitted 
\bibitem{neuh06} Neuh\"auser, R., Seifahrt, A., Roell, T., Bedalov, A., \& Mugrauer, M.\ 2006, IAU Symposium, 240, 
\bibitem{neuh05} Neuh{\"a}user, R., Guenther, E.~W., Wuchterl, G., Mugrauer, M., Bedalov, A., \& Hauschildt, P.~H.\ 2005, A\&A , 435, L13 
\bibitem{Mayor} Mayor, M., Udry, S., Naef, D., Pepe, F., Queloz, D., Santos, N.~C., \& Burnet, M.\ 2004, A\&A, 415, 391 
\bibitem{mclaughlin} McLaughlin, D.~E., 
Anderson, J., Meylan, G., Gebhardt, K., Pryor, C., Minniti, D., \& Phinney, 
S.\ 2006, ApJS, 166, 249 
\bibitem{Pravdo} Pravdo, S.~H., \& Shaklan, S.~B.\ 1996, ApJ, 465, 264 

\end{thebibliography}
\end{document}